# Towards structural softness and enhanced electromechanical responses in HfO$_2$ ferroelectrics

Binayak Mukherjee[1*], Natalya S. Fedorova[1] and Jorge Íñiguez-González[1,2,+]

[1]Luxembourg Institute of Science and Technology (LIST), Avenue des Hauts-Fourneaux 5, L-4362, Esch-sur-Alzette, Luxembourg

[2]Department of Physics and Materials Science, University of Luxembourg, 41 Rue du Brill, L-4422, Belvaux, Luxembourg

Structural softness – often characterized by unstable phonon modes and large electromechanical responses – is a hallmark of ferroelectric perovskites like BaTiO$_3$ or Pb(Ti,Zr)O$_3$. Whether HfO$_2$ ferroelectrics present any such structural softness is still a matter of debate. Here, using first principles calculations, we predict that it is possible to induce structural instabilities in hafnia. More specifically, our calculations show that in-plane epitaxial tensile strain causes a mechanical instability of the ferroelectric phase, which transforms discontinuously into an antipolar polymorph. Then, upon release of the tensile strain, the antipolar polymorph transforms back to the ferroelectric state by a soft phonon instability. We show that the softening is accompanied by enhancements in the dielectric and piezoelectric responses. While these transitions occur at high epitaxial strains for pure ferroelectric HfO$_2$, we show that the required deformations are considerably lowered in superlattices with other simple oxides, which may facilitate realizing these effects experimentally.

Hafnia (HfO$_2$) as a ferroelectric material has attracted great interest due to its compatibility with silicon substrates, which makes it ideal for device applications. However, in comparison to well-understood perovskite ferroelectrics such as BaTiO$_3$ and Pb(Ti,Zr)O$_3$, hafnia remains something of a paradox [1]. Unlike in the former, the polar phase in HfO$_2$ (orthorhombic *Pca2$_1$*, oIII for short) is not the ground state, but rather a metastable phase which can be stabilized through a combination of epitaxial strain, point defects, and kinetics. The experimentally observed ground state is instead the monoclinic phase (*P2$_1$/c*, m in the following), with high-temperature tetragonal (t) and cubic (c) phases also being observed [2].

Conventional ferroelectrics are characterized by structural softness close to the phase transition, generally associated with the reduction in frequency (softening) of a zone center optical polar phonon mode. The structural softness which accompanies this mode softening strongly affects the response coefficients (dielectric, piezoelectric, electro-optical). These properties are controlled by low frequency polar phonons, which yield large lattice responses to mechanical and electrical perturbations – as exemplified by the Curie-Weiss law for the dielectric susceptibility [3].

In a departure from the conventional ferroelectric behavior described above, evidence for such structural softness in the ferroelectric oIII phase of HfO$_2$ has been conspicuously lacking. A peak in the dielectric constant has been experimentally observed in oxygen deficient Hf$_{0.5}$Zr$_{0.5}$O$_{2-\delta}$ films close to the oIII to t phase transition temperature [4], but it remains unclear whether this effect is intrinsic or, for example, occurs due to the movement of oxygen vacancies. Nevertheless, in the closely related ZrO$_2$, molecular dynamics simulations with a machine-learned force field have shown a considerable (and purely intrinsic) enhancement in the dielectric constant of the oIII phase as the transition temperature to the paraelectric t-phase is approached [5]. However, the authors refrain from making a connection to mode softening, due to the absence of soft modes in the higher symmetry phases of $HfO_2$ and $ZrO_2$ [6,7]. Inflections in the $e_{33}$ component of the piezoresponse (which is proportional to the dielectric response $\chi_{33}$) have also been reported in oIII $HfO_2$ from first principles, as a function of epitaxial strain applied in both the (001)- [8] and (111)-planes [9]. In the former case, the authors observed that such an inflection point – which marks the start of an accelerated increase in the magnitude of $e_{33}$ – occurs just before the oIII phase loses its stability.

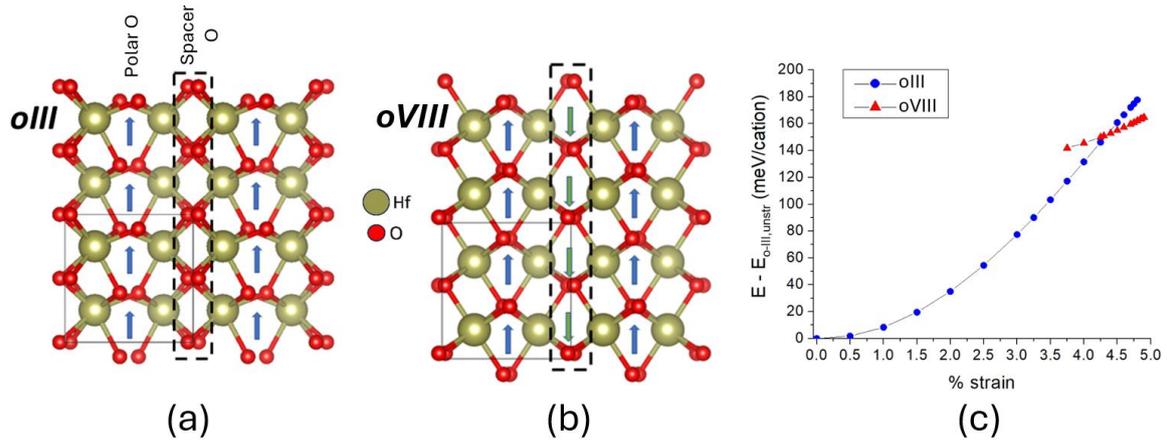

Fig 1. Structures of polar (a) and antipolar (b) HfO2, with the spacer oxygen, and the dependence of their energy on epitaxial strain (c).

In this article we use first-principles methods to further investigate this destabilization, with increasing tensile strain, of the polar oIII phase into another structure, which we identify as the closely related antipolar orthorhombic polymorph (*Pbcn*, oVIII). This antipolar polymorph, while not observed experimentally, has been discussed as a potential high-symmetry reference for the polar oIII structure [10,11]. We further explore the converse transformation of the oVIII phase into oIII as the epitaxial strain diminishes. Our results demonstrate the onset of structural softness in $HfO_2$ at the limits of phase stability, directly connected with enhancements in the electromechanical responses close to the critical point.

The crystal structures of the oIII (Fig. 1a) and oVIII (Fig. 1b) phases are closely related. The oIII structure contains two symmetry-inequivalent oxygens, one in an approximately centrosymmetric position (spacer oxygen) and one in a position with a lower-symmetry environment (polar oxygen). By contrast, in the oVIII structure all the oxygens have a low-symmetry environment similar to that of the "polar oxygens" of the oIII phase, but displaying an anti-polar pattern that yields a null polarization. Interestingly, despite being structurally similar, the cell volume of the oVIII phase is about 10% larger (see Supplementary Table 1).

To challenge the stability of the oIII phase of $HfO_2$, we apply an increasing tensile epitaxial strain in the (001)- or AB-plane in a uniform manner ($\eta_{epi} = \eta_A = \eta_B$), where [001] is the polar axis. Throughout the article $\eta_{epi}$ is always defined with respect to the lattice parameters of the bulk oIII phase. As shown in Fig. 1c, the energy of the oIII phase increases with increasing $\eta_{epi}$, before it becomes unstable around $\eta_{epi} = 4.85$ %, eventually relaxing into a strained version of the oVIII phase. If we now reduce $\eta_{epi}$ for this oVIII state, we find that below 4.5% it becomes unfavorable vis-à-vis the oIII phase at the same strain state; then, at $\eta_{epi} = 3.75\%$, the oVIII phase stops being a local minimum of the energy and the structure

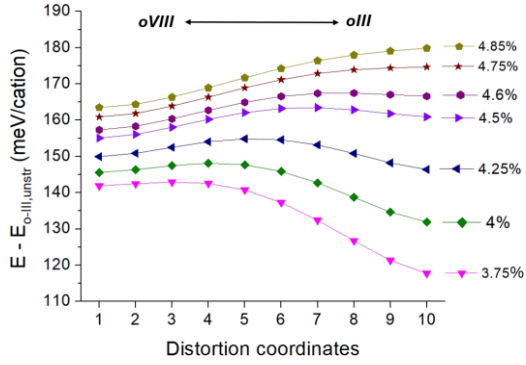

Fig 2. Energy barriers between the antipolar oVIII (left) and polar oIII (right) phases of HfO2, for a range of strain states.

evolves toward the corresponding oIII solution at the same $\eta_{epi}$.

Let us stress that, at any given $\eta_{epi}$, we consider the same in-plane lattice vectors for both the oIII and oVIII polymorphs. For the oVIII phase, compared to the fully relaxed bulk polymorph, this corresponds to having a tensile strain in one in-plane direction and a compression along the other, even at $\eta_{epi} = 0$ %. Hence, in this study we never consider bulk oVIII hafnia, which is predicted to be a

To better understand the nature of these oIII-oVIII transformations, Fig. 2 shows the energy landscape between the two polymorphs as obtained from a simple structural interpolation, and its variation as a function of epitaxial strain. We have two minima separated by an energy barrier in a wide range of $\eta_{epi}$ values. At the limits of the coexistence region, one polymorph loses its stability and transforms into the other. Hence, the evidence for some sort of structural softening is clear.

To identify the origin of the structural instabilities, we track the frequency of the Γ phonons of the oIII and oVIII phases, as well as their elastic constants, as a function of epitaxial strain. As shown in Fig. 3a, the lowest lying optical phonon in the oIII phase (blue circles) softens significantly with increasing strain. However, the oIII local minimum disappears at $\eta_{epi} = 4.85\%$ (Fig. 2), well before this mode can become unstable; hence, the destabilization of the oIII phase is not driven by an optical mode. Interestingly, this softening mode has polar symmetry, but its polarity is minimum of the energy when relaxed under zero stress [12] with no phonon instabilities (see Supplementary Fig. 1 for phonon dispersions of the bulk oIII and oVIII phases).

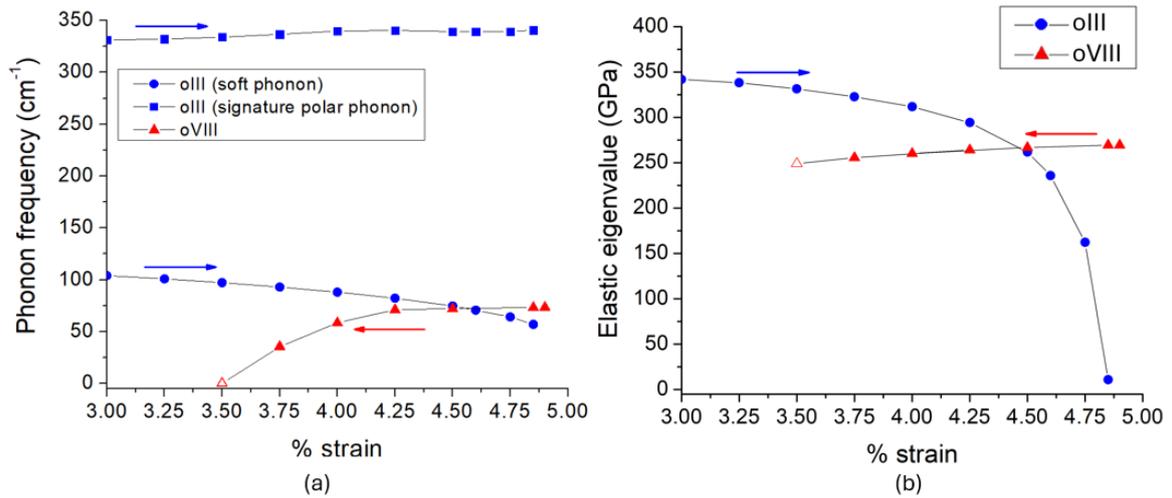

Fig 3. (a) Frequency of Γ phonons and (b) elastic eigenvalues for the oIII and oVIII phases of HfO$_2$. The open symbol 'Δ'; indicates a dynamically unstable structure.

relatively weak compared to the so called "signature" polar mode [13], a higher energy optical phonon that does not soften appreciably upon application of the tensile strain (blue squares in Fig. 3a).

What then may cause the instability of the oIII phase (clearly shown in Fig. 2) if all the optical modes are stable? As it turns out, we find that the destabilization of the oIII phase is due to a mechanical instability. Indeed, as shown in Fig. 3b (blue circles), the lowest-lying eigenvalue of the clamped elastic tensor $C_{ijkl}$ drops sharply with increasing strain, and it eventually becomes null. (See Supplementary Note 1 for details about the calculation of the substrate-clamped elastic tensor.) Thus, our calculations suggest that the strain-driven transition from the polar oIII phase to the antipolar oVIII polymorph is a proper ferroelastic one [14], while the disappearance of the polarization is a secondary effect.

the oVIII phase loses its stability via a soft polar mode that drives a proper ferroelectric transition to the oIII state. (See Supplementary Fig 1. for phonon dispersions of relaxed and strained oIII and oVIII polymorphs.)

The predicted structural transitions warrant some additional discussion. We find that the elastic instability of the oIII phase (Fig. 3a) corresponds essentially to an out-of-plane deformation of the cell. Interestingly, as we approach the instability point ($\eta_{epi} = 4.85\%$), we can observe a clear increase in the out-of-plane lattice constant of the oIII polymorph. These results, shown in Fig. 4a, are striking, as the expected behavior under a growing tensile

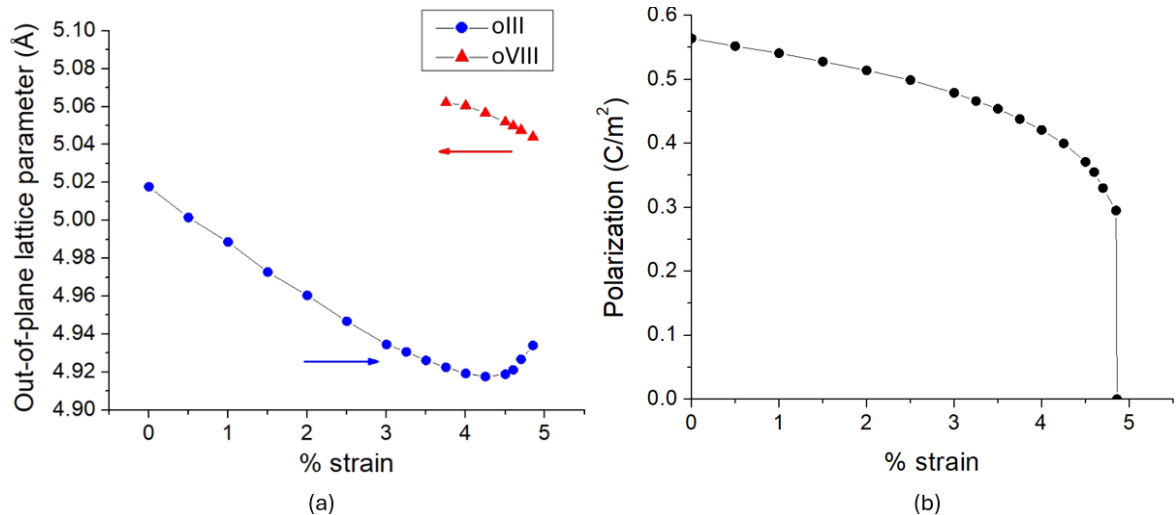

Fig 4. (a) Evolution of the out-of-plane lattice parameter of the oIII and oVIII phases with strain, showing the onset of the auxetic regime and (b) the smooth reduction of the polarization of the oIII phase with strain before transitioning discontinuously to the oVIII phase.

Turning now to the oVIII phase, we find it remains mechanically stable throughout the investigated strain range, with the lowest-lying eigenvalue of the clamped elastic tensor softening only marginally (red triangles in Fig. 3b). Instead, the frequency of the lowest lying optical Γ mode (red triangles in Fig. 3a) drops smoothly to zero and becomes unstable as $\eta_{epi}$ decreases. This soft mode is polar in nature and, when condensed, it takes the system to the oIII phase. Hence, in the particular epitaxial conditions considered in our calculations – which, as mentioned above, impose a deformation to bulk oVIII even at $\eta_{epi} = 0\%$ – strain is a shrinking of the ouf-of-plane lattice constant. Instead, close to the critical point, the oIII phase enters what we could call an "auxetic regime" that is reminiscent of previous discussions of the electromechanical response of this polymorph [15,16]. Eventually, as the oIII phase becomes unstable, the out-of-plane lattice constant starts to grow to approach its value in the oVIII polymorph. This effect can also be appreciated in the evolution of the residual epitaxial stress imposed by the substrate, as shown in Supplementary Fig. 3. Interestingly, the Z-component of polarization ($P_Z$) in the oIII phase drops smoothly with strain (Fig 4b), again evidencing how the oIII phase approaches the non-polar oVIII polymorph for increasing $\eta_{epi}$.

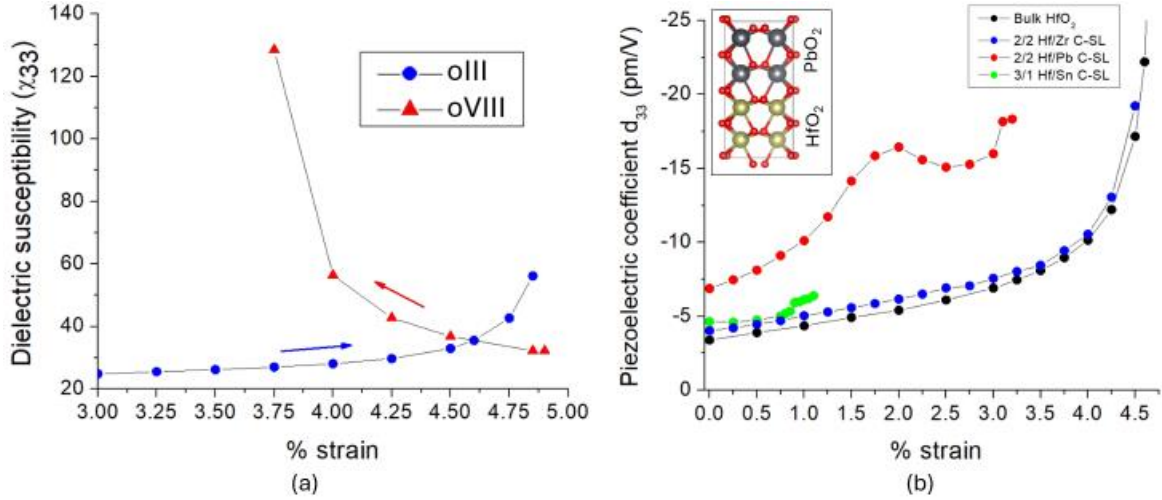

Fig 5. (a) Evolution of the dielectric susceptibility χ₃₃ of the oIII and oVIII phases of HfO₂ and (b) the clamped piezoelectric coefficient $d_{33}$ of pure oIII HfO₂ and its ferroelectric superlattices (2/2 Hf/Pb SL shown in inset).

In Supplementary Note 2 and Supplementary Fig. 4 we give further details on the transitions between the oIII and oVIII phases, showing results for the energy landscape as a function of the symmetry-adapted distortions connecting these polymorphs. These results clearly show that the polar distortion of the oIII phase is always stable, as compatible with the proper ferroelastic character of oIII to oVIII transformation. They also show how the oVIII structure progressively develops into a shallow (and then flat) minimum as a polar distortion becomes unstable.

Let us now discuss how the predicted structural softening affects the response properties of HfO₂. The dielectric and piezoelectric responses of the oIII and oVIII phases, as obtained with density functional perturbation theory calculations, are shown in Fig. 5. As regards the dielectric susceptibility $\chi_{33}$, our results show an enhancement as we approach the stability limit of both phases. In particular, $\chi_{33}$ grows very significantly in the case of the oVIII phase, displaying a clear Curie-Weiss behavior as expected from the condensation of the polar soft mode discussed above. By contrast, the lowest-lying polar modes of the oIII phase never become unstable, and the corresponding enhancement in $\chi_{33}$ is modest. Note that the behavior of $\chi_{33}$ of the oIII phase is mimicked by the closely related indirect piezoelectric coefficient $e_{33} = \partial P_3/\partial \eta_3$ (see Supplementary Fig. 5).

Even more interesting are our results for the direct piezoelectric coefficient $d_{33} = \partial P_3/\partial \sigma_3 = \sum_i (C^{-1})_{3i} e_{i3}$, where $\sigma$ is the stress, $C^{-1}$ is the inverse of the clamped elastic tensor (i.e., the clamped compliance tensor), and $e$ is the indirect piezoelectric tensor. As shown in Fig. 5b, $d_{33}$ follows a Curie-Weiss law of sorts as the oIII phase approaches its stability limit for increasing $\eta_{epi}$, reaching very high values. For example, our calculations yield $d_{33} = -45\ pm/V$ for $\eta_{epi} = 4.75\%$, an effect more than 10 times stronger than of the bulk oIII phase. Remarkably, at variance with the case of very strong piezoelectric perovskites like Pb(Zr,Ti)O₃, here the enhancement of the piezoelectric response does not rely on a polar soft mode. Indeed, in our strained oIII HfO₂, the values of $\chi_{33}$ and $e_{33}$ are relatively modest. Instead, the anomaly in $d_{33}$ stems directly from the elastic instability that yields a zero eigenvalue of the $C$ tensor (Fig. 3a). To our knowledge, this is the first example of a ferroelectric material displaying a purely elastic instability, which results in a divergence of the associated elastic compliance and piezoelectric response.

While structural softening in pure $HfO_2$ as a function of epitaxial strain is promising in principle, the actual value of strain required to achieve sizeable response enhancements is probably too large to reach experimentally. However, superlattices (SLs) of $HfO_2$ with other binary oxides have been proposed as suitable candidates to stabilize the polar oIII phase [17]. In particular, among the SLs we investigated in a recent study, the combinations Hf/Zr, Hf/Sn and Hf/Pb are found to enhance the $e_{33}$ coefficient, while the combinations Ge/Hf, Ti/Hf and Ce/Hf switch its sign. Further, the pure materials $PbO_2$ and $SnO_2$ present oVIII ground states, and no oIII local minimum, which suggests that layers of such compositions may help destabilize the oIII phase of the $HfO_2$, which our results above indicate is a good strategy to obtain an enhanced $d_{33}$ response.

We test this hypothesis by considering the behavior under tensile strain of three representative (001)-oriented short period superlattices – i.e., 2/2 for Hf/Zr and Hf/Pb, 3/1 for Hf/Sn. Here, the n/m notation refers to the thickness of the respective layers measured by the number of cation planes; for example, the inset of Fig. 5b shows the supercell used in our simulations of the 2/2 Hf/Pb SL. In all these SLs at zero epitaxial strain, both $HfO_2$ and the other oxide layer are in the oIII phase, with the polarization along the stacking direction. Our results predict that the destabilization of the oIII phase for an increasing tensile strain occurs in all the considered superlattices (Supplementary Fig. 6). While the Hf/Zr SL closely follows the behavior of pure $HfO_2$, the Hf/Pb and Hf/Sn systems become unstable at significantly smaller values of $\eta_{epi}$, 3.2% and 1.1% respectively. Fig. 5b shows the $d_{33}$ for the SLs as a function of strain. In all cases, the response grows in magnitude as we approach the stability limit; Hf/Zr matches closely the results for pure $HfO_2$, while in Hf/Pb and Hf/Sn the enhancement of $d_{33}$ is more modest. Our phonon calculations (Supplementary Fig. 7) for the Hf/Pb and Hf/Sn SLs suggest an intricate behavior with multiple low-lying modes and undulatory trends as seen in the results for $d_{33}$ of the Hf/Pb SL. We also find that, for the Hf/Pb and Hf/Sn systems, a significant contribution to the enhanced $d_{33}$ comes from the mechanical softness of the pure $SnO_2$ and $PbO_2$ compounds vis-à-vis pure $HfO_2$, which is elastically rather stiff. Thus, while complex and not fully conclusive, our results do suggest that suitable SLs may offer a way to further tune the oIII phase of $HfO_2$, driving it towards structural softness and enhanced responses.

In conclusion, our simulations predict that epitaxial strain can be a route to obtain structural softness in hafnia, either by creating an elastic instability (as predicted to occur in the usual ferroelectric oIII phase) or by inducing soft polar modes (as predicted to occur in a strain-stabilized anti-polar oVIII polymorph). Said structural softness results in associated Curie-Weiss-like behaviors for the piezoelectric and dielectric response, and in particular we predict that $d_{33}$ values of the order of -45 pm/V may be attainable for strains of the order of 4.75%. Further, our calculations provide evidence that working with $HfO_2$-based superlattices -- involving other simple oxides like $PbO_2$ or $SnO_2$ – may allow us to achieve enhanced functional responses at moderate tensile strains (∼ 1-2%). We hope these results will further our understanding of hafnia ferroelectrics and the conditions in which they may present structural softness and competitive electromechanical properties.

*Computational details*

The density functional theory calculations were performed using the plane-wave basis set as implemented in the Vienna ab-initio simulation package (VASP) [18,19]. The Perdew-Burke-Ernzerhof (PBE) form of the generalized gradient approximation (GGA) with the PBEsol modification is employed to approximate the exchange-correlation functional [20]. The plane-waves are expanded up to a cutoff of 600 eV. The following valence states are explicitly considered for the different elements: O – $2s^2$, $2p^4$; Hf – $5p^6$, $5d^2$, $6s^2$; Zr – $4s^2$, $4p^6$, $4d^2$, $5s^2$, Sn – $4d^{10}$, $5s^2$, $5p^2$; Pb – $5d^{10}$, $6s^2$, $6p^2$. For bulk structures, a 4x4x4 k-mesh was used to sample the Brillouin zone, with a proportional reduction to 4x4x2 for the superlattices. The Hellmann-Feynman forces on each atom were relaxed below 0.01 eV/Å for all structures. The polarization is calculated as the product of the nominal charges (+4 for cations, -2 for oxygen), and the displacements of the ions with respect to the high symmetry cubic fluorite parent structure (Fm-3m). The symmetry adapted modes are obtained with the ISODISTORT package [21,22]. VASPKIT [23] is used for post-processing while VESTA [24] is used to visualize the structures.

*Acknowledgements*

We would like to thank Hugo Aramberri (LIST) for many useful discussions. Work supported by the Luxembourg National Research Fund though grant Nos. INTER/NOW/20/15079143/TRICOLOR (B.M., J.Ì.-G.) and 21/MS/15799044/FERRODYNAMICS (NSF).

*Authors to whom correspondence should be addressed*

*binayak.mukherjee@list.lu

†jorge.iniguez@list.lu

# Towards structural softness and enhanced electromechanical responses in HfO$_2$ ferroelectrics (Supplementary information)

Binayak Mukherjee[1,*], Natalya S. Fedorova[1] and Jorge Íñiguez-González[1,2,+]

[1]Luxembourg Institute of Science and Technology (LIST), Avenue des Hauts-Fourneaux 5, L-4362, Esch-sur-Alzette, Luxembourg

[2]Department of Physics and Materials Science, University of Luxembourg, 41 Rue du Brill, L-4422, Belvaux, Luxembourg

Supplementary Table 1

Calculated lattice parameters of fully relaxed (bulk) and strained oIII and oVIII phases of HfO$_2$. (strain is always defined with respect to fully relaxed oIII lattice parameters).

| Strain-state | oIII | | | oVIII | | |
|---|---|---|---|---|---|---|
| | a | b | c | a | b | c |
| Bulk | 5.203 | 4.996 | 5.018 | 5.690 | 4.882 | 5.186 |
| 0.50% | 5.230 | 5.021 | 5.002 | - | - | - |
| 1.00% | 5.255 | 5.046 | 4.989 | - | - | - |
| 1.50% | 5.281 | 5.071 | 4.973 | - | - | - |
| 2.00% | 5.307 | 5.096 | 4.960 | - | - | - |
| 2.50% | 5.333 | 5.121 | 4.947 | - | - | - |
| 3.0% | 5.359 | 5.146 | 4.934 | - | - | - |
| 3.25% | 5.372 | 5.158 | 4.931 | - | - | - |
| 3.50% | 5.385 | 5.171 | 4.926 | 5.385 | 5.171 | 5.066 |
| 3.75% | 5.398 | 5.183 | 4.922 | 5.398 | 5.183 | 5.062 |
| 4.00% | 5.411 | 5.196 | 4.919 | 5.411 | 5.196 | 5.061 |
| 4.25% | 5.424 | 5.208 | 4.918 | 5.424 | 5.208 | 5.057 |
| 4.50% | 5.437 | 5.221 | 4.919 | 5.437 | 5.221 | 5.052 |
| 4.75% | 5.450 | 5.233 | 4.928 | 5.450 | 5.233 | 5.047 |
| 4.80% | 5.452 | 5.236 | 4.924 | 5.453 | 5.236 | 5.045 |
| 4.85% | 5.455 | 5.238 | 4.934 | 5.455 | 5.238 | 5.044 |

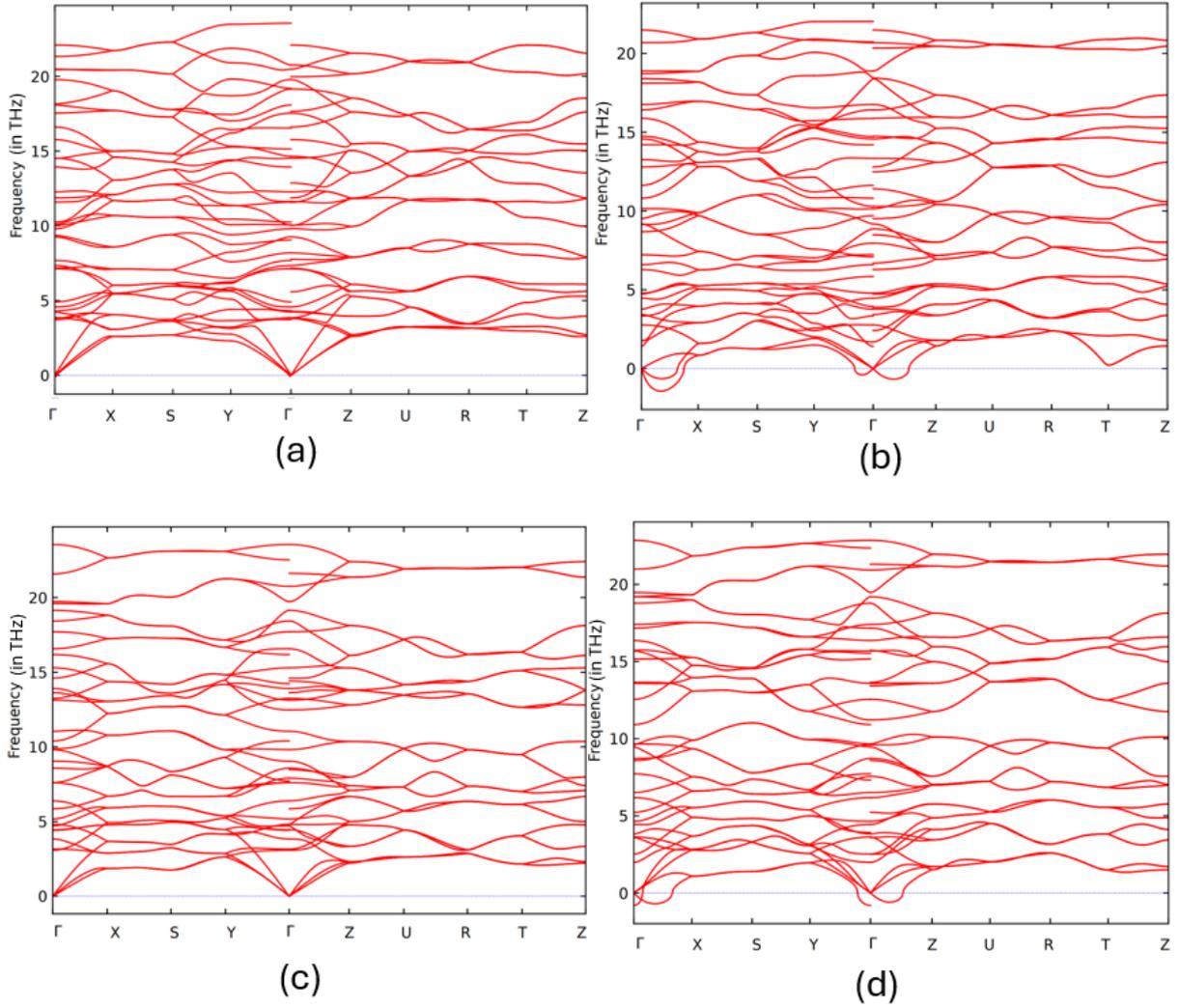

Supplementary Figure1. Phonons for (a) the fully relaxed oIII structure; (b) the mechanically unstable oIII structure ($\eta_{epi} = 4.85\%$), (c) the fully relaxed oVIII structure; and (d) the dynamically unstable oVIII structure ($\eta_{epi} = 3.5\%$)

**Supplementary Note 1:** Epitaxial clamping

When simulating epitaxial strain, it should be noted that the clamping of the film to the substrate dampens the mechanical response of the material in the plane. Accordingly, a correction has to be made to the elastic moduli tensor as calculated from first principles, since other than constraining lattice parameters, no further effects from the substrate are included in the calculations. With the 4$^{th}$-rank tensor $C_{ijkl}$ written in Voigt notation as a 6x6 matrix ($C_{ij}$), we set $C_{11}$, $C_{22}$ and $C_{66}$ to very large values, corresponding to infinite stiffness in the AB-plane. This becomes particularly relevant with respect to the mechanical instability in the oIII phase, which would be predicted to develop at a lower strain in the absence of this clamping correction. For example, at $\eta_{epi} = 4.85\%$, just before instability, the as-computed tensor is given by,

$$C_{ij} = \begin{pmatrix} -1556 & -787 & -839 & 0 & 0 & 0 \\ -787 & -52 & -318 & 0 & 0 & 0 \\ -839 & -318 & -142 & 0 & 0 & 0 \\ 0 & 0 & 0 & 47.8 & 0 & 0 \\ 0 & 0 & 0 & 0 & 77.7 & 0 \\ 0 & 0 & 0 & 0 & 0 & -190.6 \end{pmatrix} GPa$$

which shows significant softness and has negative eigenvalues corresponding to in-plane deformations. However, applying the clamping corrections we have:

$$C_{ij}^{cl} = \begin{pmatrix} 10^{10} & -787 & -839 & 0 & 0 & 0 \\ -787 & 10^{10} & -318 & 0 & 0 & 0 \\ -839 & -318 & -142 & 0 & 0 & 0 \\ 0 & 0 & 0 & 47.8 & 0 & 0 \\ 0 & 0 & 0 & 0 & 77.7 & 0 \\ 0 & 0 & 0 & 0 & 0 & 10^{10} \end{pmatrix} GPa$$

which removes the in-plane instabilities.

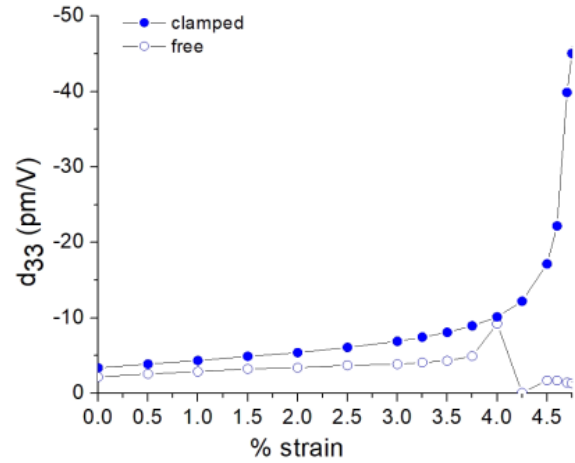

Supplementary Figure 2. Free and clamped piezoelectric coefficient $d_{33}$ as a function of strain.

This is also reflected in the difference between the uncorrected piezoelectric coefficient $d_{33}^{free}$ and its corrected counterpart $d_{33}$ – as seen in the figure above. With increasing strain, the $d_{33}^{free}$ fluctuates and becomes ill-behaved (as the *uncorrected* elastic eigenvalues becomes unstable), while the $d_{33}$ diverges quite smoothly as the instability in the *clamped* elastic eigenvalues is approached.

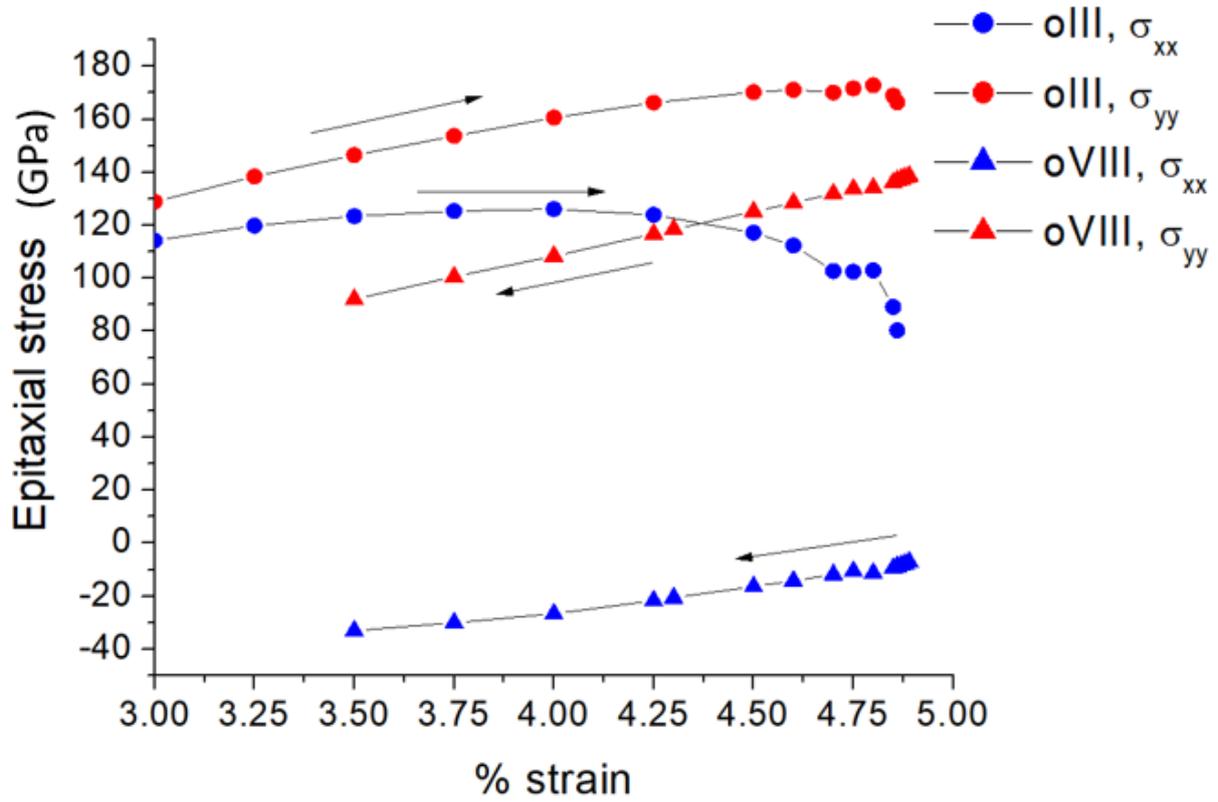

Supplementary Figure 3. In-plane residual epitaxial stress in the oIII and oVIII phases. As the oIII phase approaches its stability limit at $\eta_{epi} = 4.85\%$, the corresponding in-plane stresses decrease in magnitude and approach the relatively small values computed for the oVIII phase.

**Supplementary Note 2:** Transitions between oVIII and oIII

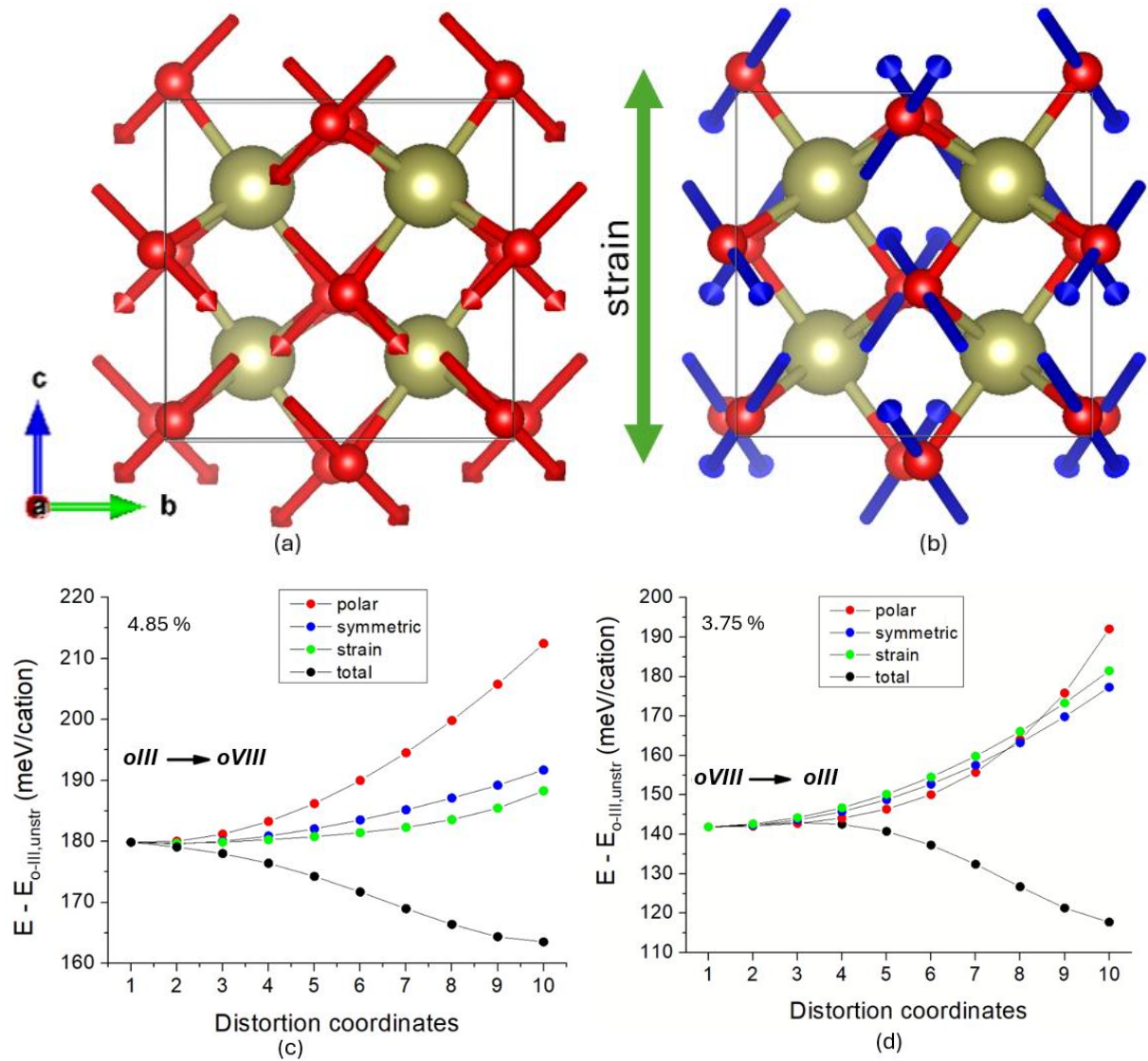

Supplementary Figure 4. (a) Polar (red arrows) and (b) symmetric distortions (blue arrows, green arrow for strain) of the oVIII structure connecting it to the oIII phase. Energy variation as we distort the material following modes of different symmetries, when going from oIII to oVIII at $\eta_{epi} = 4.85\%$(c) and from oVIII to oIII at $\eta_{epi} = 3.75\%$ (d).

Based on the analysis of symmetry adapted modes, the high symmetry oVIII phase can be distorted into the polar oIII phase through a combination of displacements: (i) a polar distortion involving *in essence* the polar and spacer oxygens (Suppl Figure 4a); (ii) oxygen displacements and strains that respect the symmetry of the oVIII phase (Suppl Figure 4b). When comparing structures in identical strain states, in-plane strain modes are naturally zero. By plotting the energy of these displacements separately as a function of the distortion amplitude, we can gain insight into the energy barriers between the two phases (for total barriers see Fig 2 in main text), allowing a quantification of the energy costs of the individual distortions.

We first connect the two phases at $\eta_{epi} = 4.85\%$, just before the polar oIII phase becomes mechanically unstable (Suppl Fig 4c). Since the oIII phase is the lower symmetry polymorph, we remove the aforementioned distortions from the oIII phase to obtain oVIII. Doing so individually leads to increases in the energy of the system, which is brought down by the combination of the distortions, resulting in the lower energy oVIII phase. Crucially, the polar distortion has a distinctly higher energy cost, manifested in a steeper slope in the energy. By contrast, the stain distortion has a very flat energy landscape associated with it, consistent with the fact that a purely elastic instability is about to develop.

Then, at $\eta_{epi} = 3.75\%$, just before the oVIII phase develops a dynamical instability, the polar distortion that leads to the oIII phase has a very flat energy variation associated to it, hinting at the imminent occurrence of a soft mode instability (Suppl Fig 4d).

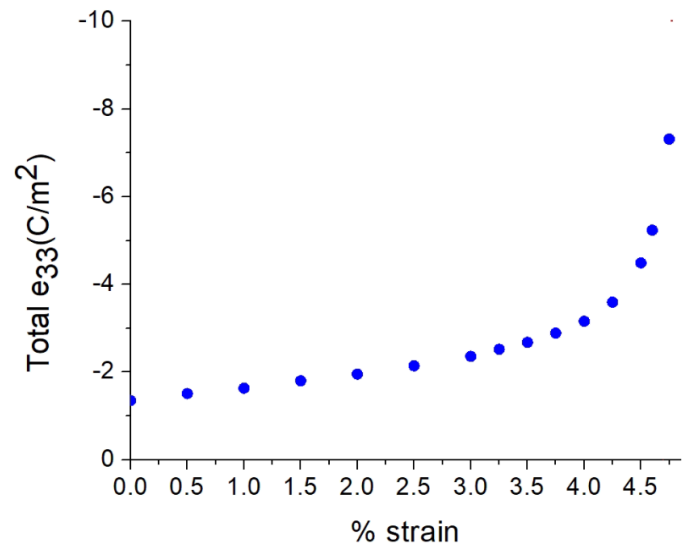

Supplementary Figure 5. Piezoelectric response $e_{33}$ of oIII $HfO_2$ as a function of strain.

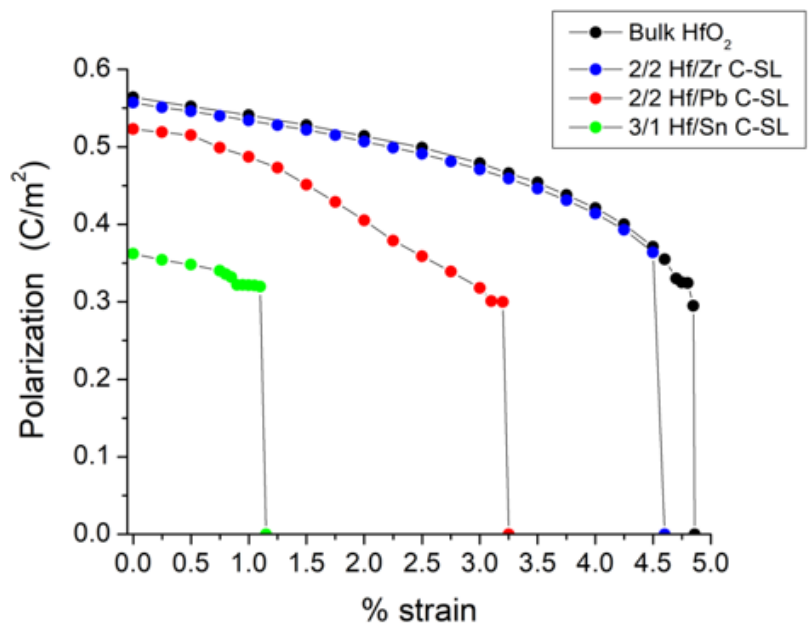

Supplementary Figure 6. Z-component of polarization of bulk $HfO_2$, and the superlattices Hf/Zr, Sn/Hf, and Pb/Hf, showing the respective destabilization of the polar oIII phase as a function of strain.

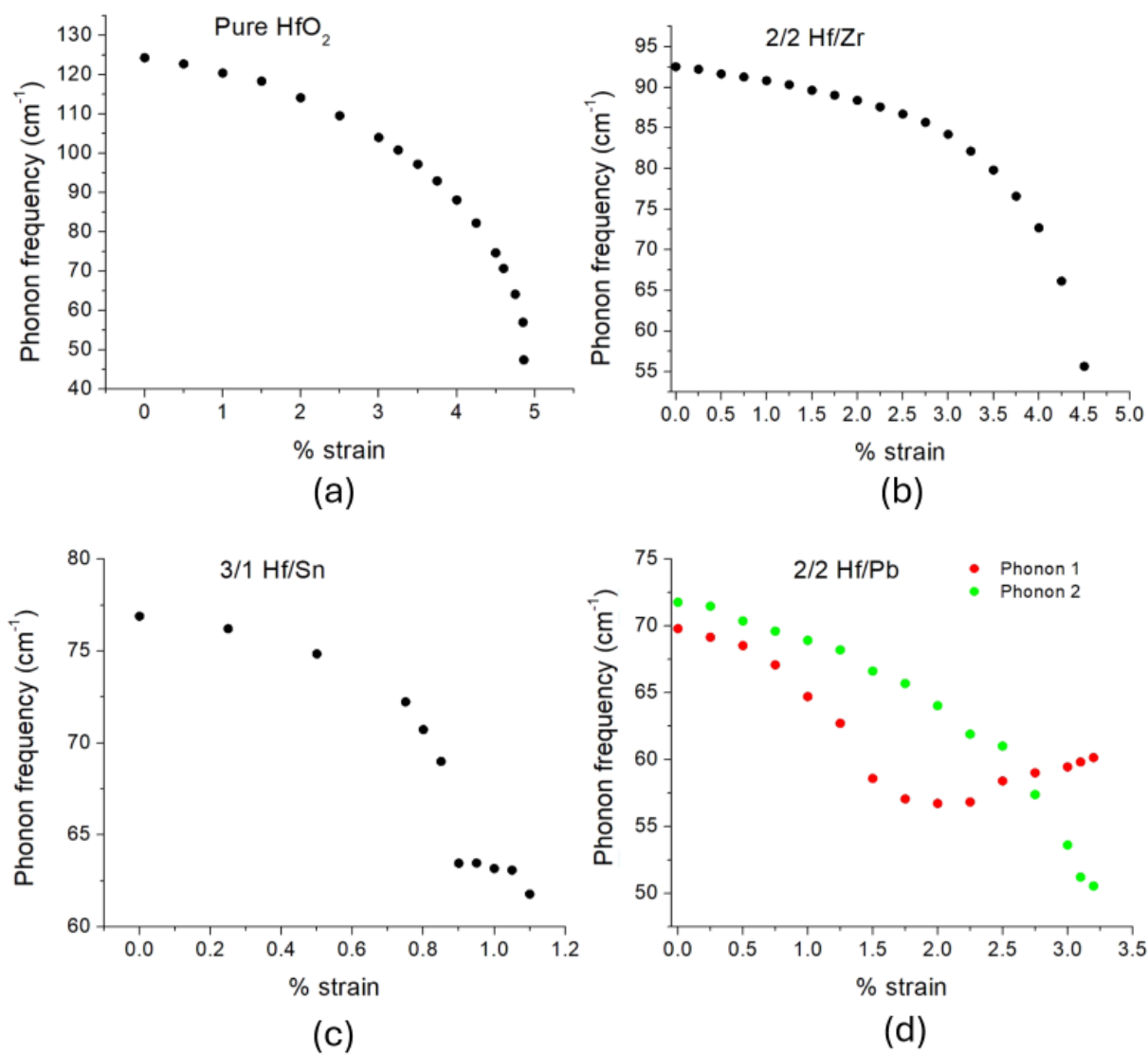

Supplementary Figure 7. Lowest lying optical phonons at Γ in (a) bulk HfO2, and the superlattices (b) Hf/Zr; (c) Sn/Hf, and (d) Pb/Hf. The frequency scale has been adjusted in each case to highlight the softening of the phonons.